\documentclass{elsart}
\usepackage{graphicx}    
\journal{Astroparticle Physics}           
\begin{document}
\begin{frontmatter}

\title{The puzzling MILAGRO hot spots}

\author{L. O'C. Drury},
\author {F. A. Aharonian}
\address{
Dublin Institute for Advanced Studies,
31 Fitzwilliam Place,
Dublin 2, Ireland.
}

\begin{abstract}
We discuss the reported detection by the MILAGRO experiment of localised hot spots in the cosmic ray arrival distribution and the difficulty of interpreting these observations. A model based on secondary neutron production in the heliotail is shown to fail. An alternative model based on loss-cone leakage through a magnetic trap from a local source region is proposed.
\end{abstract}
\end{frontmatter}

\section{Introduction}

The MILAGRO experiment \cite{MILAGRO} has reported the surprising detection of two localised ``hot spots'' in the cosmic ray arrival distribution the stronger and more localised of which (region A) coincides with the local heliotail direction.  While the intensity is quite low (fractional excess of order $6\times 10^{-4}$ in region A) the observation of what appear to be essentially  well-collimated beams of hadrons (with energy about 10 TeV or more) coming from very specific directions is hard to accommodate within standard models of cosmic ray propagation. 
Although the  observed tiny small-scale 
anisotropy could be most naturally  explained as being due to gamma-rays,   the MILAGRO group  claims that an electromagnetic origin of the signal is excluded with high confidence.  This leaves protons
and nuclei, and, to a lesser extent, neutrons  as particles which could cause the 
observed anisotropy.
However as noted in the MILAGRO discovery paper protons at this energy only have gyroradii of order $0.005 \,\rm pc\approx 10^{3}\,AU$ in the estimated local interstellar magnetic field of $2\,\mu\rm G$ and the decay length of neutrons is only $0.1 \, \rm pc$.   
These simple, but robust,  estimates indicate that any
interpretation of this result will require nonstandard ideas and approaches. 

That there should be some structure in the arrival direction distribution is very natural.  Apart from large scale gradients related to the bulk Galactic propagation, it would be very strange if there were not some signature of nearby recent cosmic ray source events (such as local supernovae) as discussed by, for example, Erlykin and Wolfendale \cite{EW}.   In this context it has recently been proposed by Salvati and Sacco \cite{SS} that what MILAGRO sees is a signature of the supernova explosion that resulted in the formation of the Geminga pulsar.   But one would expect all these processes to give rise to low-order multipole components, and indeed predominantly a dipole component, in the angular distribution.  The peculiarity of the MILAGRO result is that it shows discrete features on small angular scales, and indeed not just  random power in high-order multipoles, but quite discrete localised features.  The data analysis technique used by MILAGRO does correspond to a high-pass filter which emphasises these local hot-spots, but they cannot be dismissed as mere artifacts of the analysis.  

\section{Neutron source models}

Motivated largely by the coincidence between the heliotail direction and region A, we first consider the possibility that region A might be explicable in terms of secondary neutron production in the gravitationally focused tail of ISM material that forms  downstream of the Sun as it moves through the local ISM.  Unfortunately, while the target has exactly the right size and location, the density is much too low to account for the observed signal (quite apart from the fact that this model would fail to account for region B even were it to work for region A). 

\subsection{Neutron induced enhancement}

Observing into a narrow cone of solid angle $d\Omega$, a thin gas target of number density $n_{\rm T}$ at distance $r$ bathed in an isotropic and uniform flux of cosmic rays of number density $n_{\rm CR}$ will produce a flux of neutrons
\begin{equation}
{r^2\,dr\,d\Omega\,n_{\rm T} \sigma_{\rm pN} c n_{\rm CR} \over 4 \pi r^2},
\end{equation}
where $\sigma_{\rm pN}$ is the cross-section for converting a proton to a neutron,
either via the delta resonance or by directly knocking a neutron out of an atomic nucleus.
The same flux of cosmic rays will produce a background rate within the cone of 
\begin{equation}
{d\Omega\over 4\pi} c n_{\rm CR}.
\end{equation}
Thus the relative enhancement of the signal will be simply
\begin{equation}
\int n_{\rm T} \sigma_{\rm pN} \, dr
\end{equation}
and the question is whether we can get enough target material into the cone to give an enhancement factor of $10^{-3}$.  
The ratio of the flux of secondary neutrons to cosmic rays  can be estimated as  \cite{DAV} 
\begin{equation}
\frac{f_{\rm n}}{f_{\rm CR}}=
\left<mx\right>_{\rm n}^\alpha \, \sigma_{\rm pp} \eta N_{\rm H} \approx 2\times 10^{-27} N_{\rm H} 
\end{equation}
where the first factor is the so-called spectrum weighted moment for neutron production, 
$\sigma_{\rm pp}$ is the inelastic proton-proton cross-section and $\eta\approx 1.5$ is a correction factor to account for the presence of heavy nuclei.  We note that while the production energy spectrum of the neutrons will be almost exactly the same as that of the ambient cosmic rays, 
the energy 
dependent ($\propto E$) decay time will distort the spectrum and make 
it harder at certain energies, if the source is located at the ``right'' distance from us.  For the estimated energy 
of the shower initiated particles of $\sim 10 \ \rm TeV$, the source of neutrons 
should not be much further (otherwise the flux would be dramatically suppressed)
and  not much closer (otherwise the effect of spectral hardening would disappear) 
than  $d=t_{\rm decay}(10 \ \rm TeV) c = \gamma_n t_n c\approx 0.1 \, \rm pc$ where $\gamma_n$ is the Lorentz time-dilation factor of the neutron and $t_n$ is the free neutron decay time.

\subsection {Gravitational focusing of ISM material}

It has long been known that the gravitational field of the sun can focus interstellar neutral atoms (and dust grains) as they flow past the Sun leading to the formation of a density caustic on the downstream symmetry axis.   The existence of such an effect has been directly confirmed by observations with the Ulysses spacecraft and indirectly by optical back-scatter observations \cite{Ulysses, Lyman}.  Detailed analytical and numerical studies of this effect were first published by Fahr and co-workers back in the early 1970s \cite{F71,BF70}.

For our purposes a simple impulse approximation is adequate.  A neutral particle moving past the Sun at impact parameter $b$ and velocity $v$ experiences an acceleration of order $GM/b^2$ lasting for a time of order $2b/v$.  It thus acquired a radial inward drift velocity of order 
\begin{equation}
u = {2 b\over v} {G M \over b^2} = {2 G M\over v b}
\end{equation}
and reaches the symmetry axis after a time $b/u = v b^2/2GM$ and at a downstream distance of 
\begin{equation}
L = {v^2 b^2 \over 2 G M}.
\end{equation}
Clearly in the approximation of cold particles the density becomes infinite on the axis with a sharp cusp, but in reality the effect will be blurred by the initial velocity dispersion of the particles and in any case we can only observe with finite angular resolution. 

Let us therefore consider observing into a narrow cone of opening angle $\vartheta$ and consider first the case where the gravitational deflection is greater than $\vartheta$.
At a focusing distance downstream of $L$ the cone has a radius of $\vartheta L$ and the flux of deflected particles entering the approximately cylindrical section between $L$ and $L+dL$ is  $n_0 v 2\pi b db$ (where $b$ and $L$ are related as above) and the particles spend a time of order $2\vartheta L/u$ in the cone.  The mean number density in the cone at distance $L$ is thus
\begin{equation}
n_0 v 2 \pi b db\, {2\vartheta L\over u}\left(\pi \vartheta^2 L^2 dL\right)^{-1} = {4 n_0\over\vartheta} {b\over L}
\end{equation}
for $\vartheta < b/L$.  Once the deflection becomes comparable to the opening angle of the cone this approximation breaks down; there is still a focused component on axis, but it becomes less and less important relative to the directly inflowing material and the mean density in the cone tends asymptotically back to the ambient value $n_0$.  That the enhancement factor is about 4 when the deflection angle equals the cone opening angle is intuitively obvious because in this case the collecting radius is precisely twice the radius of the circular cross-section of the cone into which the particles are forced.

The peak column density is thus of order $4 n_0 L$ where $L$ is determined by the condition that $b\approx \vartheta L$.  Taking $\vartheta \approx 10^{-2}$  for an opening angle of about a degree and substituting values appropriate for the sun \cite{HeReview} ($v\approx 3\times 10^6 \,\rm cm\,s^{-1}$, $2GM/c^2 = 3\,\rm km$) we get a maximum value of $b$ of $3\times 10^{15}\,\rm cm$ and thus $L = 3\times 10^{17}\,\rm cm = 0.1 \,\rm pc$.  In fact the focusing will be probably be disrupted by the gravitational field of other near-by stars (at distances of a few parsecs) on scales any larger than this.

\subsection{Flux estimates}
We thus rather naturally produce a thin gas target at the appropriate distance of $0.1 \, \rm pc$, which can fit the location of region A and the harder spectrum, but unfortunately the effective enhancement of the ambient density by only a factor of four is far too weak to account for the observed signal, the density of the local interstellar medium around the sun being quite low \cite{HeReview}  with a rather accurately measured Helium number density of $0.015 \,\rm cm^{-3}$ corresponding to $n_0\approx 0.15\,\rm cm^{-3}$ and thus a column density of $1.8\times 10^{17} \rm cm^{-2}$ giving a mere $4\times10^{-10}$ enhancement rather than the $10^{-3}$ observed.

More generally we can conclude that no local neutron source model based purely on secondary production is viable because the required target density is so high, at some $10^6$ times the local density that it should have been easily seen by astronomers studying the local bubble. 

\section{Proton and ion source models}

If not neutrons, let us now consider the possibility that the signal might be due to protons
(or more generally heavy ions). Even supposing the Earth to be magnetically connected to a remote accelerator, and for ions to travel rather freely along this field, one would naively expect them to arrive with a broad distribution of pitch angles relative to the field and it is thus hard to explain the narrow angular scale of region A (the angular distribution of arrival directions is quite compact with a size of only a few degrees, whereas a gyrating streaming proton distribution should fill virtually the entire forward hemisphere of $2\pi$ steradians).  

This objection applies {\it a fortiori} to the model in \cite{SS} where the interstellar propagation is assumed to be strong-scattering diffusion in the Bohm limit.   On this model the anisotropy signal would be of dipole form with significant back-scattering of particles.  In fact, while Bohm scaling for the scattering in the immediate vicinity of the shock is quite plausible and indeed seems required, it would be extraordinary for interstellar propagation to be described in this way and it would completely contradict the conventional picture of interstellar magnetic fields and cosmic ray propagation.

Similar considerations apply to any local accelerator of protons using stochastic acceleration (first or second order fermi) all of which produce essentially isotropic particle distributions.  Magnetic reconnection can produce tightly collimated beams of particles along the magnetic null line, but it is unlikely that the fields can be so well organised over the scales required in the solar neighbourhood.  In any case there is the fundamental problem that any such local acceleration process is constrained to the Hillas limit on attainable particle rigidity of $B V L$ where $B$ is the magnetic field strength, $V$ a characteristic velocity and $L$ a characteristic length. Taking optimistic values of a $2\,\mu \rm G$ field, a length scale of $0.1\,\rm pc$ and a velocity of $30\,\rm km\, s^{-1}$ (the order of magnitude of the solar velocity with respect to the LISM) gives a maximum particle rigidity of $20\,\rm GV$, well short of the energies at which the signal is seen.   And of course if the source is less than a gyroradius away, by definition the accelerator cannot use magnetic confinement anyway and we would have to find an electrostatic acceleration potential of $10^{13}\,\rm V$ in the local neighbourhood!  In this context we note that Salvati and Sacco \cite{SS} also give an energy budget argument against a local heliospheric origin which we find compelling.

Because the essential problem is the small gyroradius one could try using particles with a larger mass to charge ratio.  In theory one could consider singly charged heavy ions (similar to the anomalous cosmic rays observed at much lower energies), but even if heavy ions could be accelerated while retaining an electron cloud, the electron stripping cross-sections are so large at these energies that any such particles would almost instantaneously become fully ionised during propagation.  In any case, even assuming the signal to be singly charged heavy ions would only gain a factor of one to two decades and thus still leave the putative accelerator closer than the nearest stars.  Charged dust grains can only be slightly accelerated before frictional and sputtering lossses intervene \cite{EDM}
and in any case on interaction with the atmosphere are not expected to produce the penetrating showers seen by MILAGRO.

One possible way out would be to suppose that we happened to be connected to a transient event and only the fastest particles (moving essentially straight along the field with little or no gyration) have reached us, but this seems very contrived (although it does make the interesting prediction that the spot will brighten and broaden out on a time scale of the light travel time to the event which could be as short as a few years).  

A more plausible alternative is to invoke relatively scatter-free propagation with magnetic moment conservation, and to suppose that the field line connecting us to the source has come from a region of high field into a much lower field region (a rather plausible scenario).  Adiabatic conservation of the magnetic moment then implies that the particles momentum component perpendicular to the field also has to drop in such a way that $p_\perp^2/B$ is constant and thus the particles can be focused into a beam parallel to the field.  Another way to think of this is in terms of Liouville's theorem on the conservation of phase-space density.  To compensate for the reduction in angular distribution from $2\pi$ steradians down to the observed $0.3$ steradians of regions A and B, the cross-section of the magnetic flux tube has to expand by an amount $2\pi/0.3\approx 20$ which seems quite plausible.

Of course if there is a significant flux of such particles this is a highly unstable velocity distribution and resonant modes will be excited which scatter the particles and attempt to restore isotropy, but for a  sufficiently weak beam this effect will not be a problem 
and the resonant modes can be damped by, among other possibilities, Landau damping on the bulk cosmic rays and ion-neutral friction \cite{VC, KP} (and we are only talking about a beam of order $10^{-3}$ relative to the background).   In addition, if the field lines continue to diverge, the adiabatic focussing effect can at least partially compensate for pitch-angle scattering as long as the characteristic length of the field variation is comparable to the scattering mean free path.  On balance this seems the least unlikely explanation of the results. A  field configuration departing from strict axisymmetry could allow the loss-cone to be defocused into the two regions observed by MILAGRO, but the alignment with the heliotail of region A would have to be largely coincidental on this explanation. However it is possible that draping of the interstellar field along the heliosheath might help and this may also account for the fact that, as noted by Abdo et al \cite{MILAGRO}, the directions of regions A and B do not seem to lie along what is thought to be the local interstellar field direction.  But it is not clear that the local field is really know at the scales of interest here; to quote
Leroy \cite{JLL} ``The interstellar magnetic field revealed by the nearest polarized stars is certainly not a smooth, uniform field along the Galactic equator''.

\section{Conclusion}

\begin{figure}[htbp] 
   \centering
   \includegraphics[width=3in,angle=270]{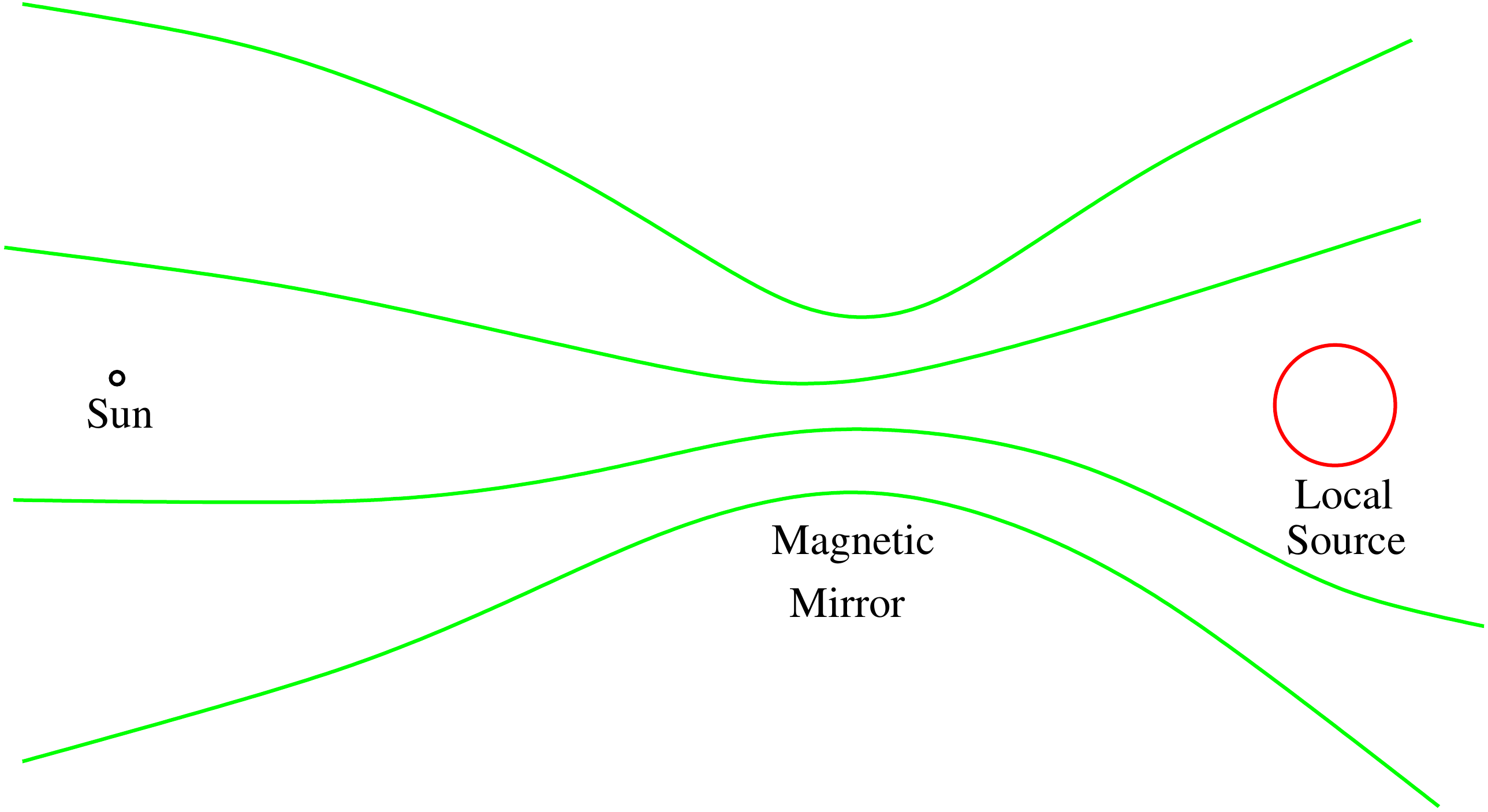} 
   \caption{A cartoon representation of the suggested model.  The magnetic field near the sun is linked through a strong field region where most particles are mirrored to a region containing a slightly higher amplitude and harder spectrum of cosmic rays, presumably as a result of a local source event such as a supernova explosion.  Only particles with pitch angles closely aligned to the field can penetrate through the mirror and emerge as a field-aligned beam on the other side (another possibility would be to locate the source inside the strong field region itself rather than behind the trap).}
   \label{trap}
\end{figure}

The MILAGRO observations of localised small-scale structures in the arrival direction distributions of cosmic rays represent a real challenge to our understanding of cosmic ray origin and propagation.  It would appear that the only viable explanation involves relatively scatter-free focused transport along a diverging magnetic field connecting the solar neighbourhood to a magnetic mirror where the field increases by about a factor of $20$ and where the cosmic ray intensity on the other side of the mirror is slightly higher in amplitude and harder in spectrum.  The particles in regions A (and presumably B?) are then those particles in the so-called ``loss cone'' which penetrate through the trap.  The bulk of the cosmic rays we see have been mirrored off the trap and reflect the general population in the solar neighbourhood.  This is  illustrated in cartoon form in Fig.~\ref{trap}.  We note in passing that relative motion between the sun and the trap (presumably a field concentration associated with a nearby molecular cloud) could also induce a large-scale signal in the anisotropy data. Further work, which must also address the question of the large scale anisotropy seen by MILAGRO as well as by the Tibet Air Shower Array and the Super-Kamiokand experiments \cite{Tibet, SuperK}, is clearly required to understand these fascinating results.

Finally, we note that while this model allows the small angular scale features to be due to a source located much further away than one proton gyroradius, it is still advantageous to have the source relatively near-by, say at less than $100\,\rm pc$.  In fact as estimated by a number of authors, such a single nearby source with a total energy release of about $10^{50}\,\rm erg$ can significantly contribute to the observed cosmic ray flux, see
for example references \cite{AAV, EW, KKYN}, and in fact there is no doubt that there must have been at least one recent nearby cosmic TeVatron from the detection of cosmic ray electrons at $E\ge 1\,\rm TeV$ \cite{Nishimura}.  Interestingly a detectable cosmic ray anisotropy from such a source has been predicted \cite{PO} and it is likely that the LAT on GLAST will be able to provide detailed studies of such TeV electrons \cite{MOM}.  Without wishing to claim any particular event as the source, the suggestion that the supernova explosion which resulted in the formation of the Geminga pulsar may have been sufficiently close and recent to leave detectable traces we find quite plausible and in this we agree with Salvati and Sacco \cite{SS}.  Other candidates, as pointed out by Kobayashi et al \cite{KKYN} who discuss this extensively, are the Vela, Cygnus loop and Monogem remnants.

\section{Acknowledgments}

LD is grateful to Brenda Dingus for drawing his attention to these remarkable observations at the Bacelona CTA meeting and for subsequent discussions.  
LD and FA thank Gus Sinnis and the MILAGRO collaboration for discussing their results with us prior to publication and John Learned for helpful comments.

\end{document}